# Band structure of new layered 17K superconductor $Sr_4Sc_2Fe_2P_2O_6$ in comparison with hypothetical $Sr_4Sc_2Fe_2As_2O_6$

**I. R. Shein* and A. L. Ivanovskii**

*Institute of Solid State Chemistry, Ural Division, Russian Academy of Sciences, Pervomaiskaya St., 91, Yekaterinburg, 620990 Russia*
*e-mail:* shein@ihim.uran.ru

**Abstract**

**The results of the *ab initio* FLAPW-GGA calculations of the band structure of the newly synthesized tetragonal (space group *P*4/n*mm*) layered iron phosphide-oxide: 17K superconductor $Sr_4Sc_2Fe_2P_2O_6$ are presented. For $Sr_4Sc_2Fe_2P_2O_6$ the optimized structural data, the energy bands, total and partial densities of states, Fermi surface topology, low-temperature electron specific heat and molar Pauli paramagnetic susceptibility have been determined and discussed in comparison with hypothetical isostructural iron arsenide-oxide phase $Sr_4Sc_2Fe_2As_2O_6$ and related layered FeAs and FeP superconductors.**

* Corresponding author.
*E-mail address:* shein@ihim.uran.ru (I.R. Shein).

## Introduction

The recent discovery of superconductivity at $T_c$'s up to 55-56 K in the layered iron-pnictide systems [1-5] has sparked enormous interest in this class of materials and has led to an intensive search for related superconductors (SCs). In result at least five groups of such materials have been discovered today, namely: (i). the group of oxygen-containing "1111" phases based on the layered arsenide oxides *Ln*FeAsO (*Ln* = La, Ce, … Gd, Tb, Dy); (ii). the related oxygen-free "1111" phases (for example *Ln*FeAsF); (iii). the group of three-component layered "122" SCs with the basis phases $AFe_2As_2$($A$ = Sr, Ba); (iv). so-called "111" SCs (LiFeAs and NaFeAs.) and (v) layered phases in binary Fe–Se(Te) systems. Besides, a lot of related systems with various pnictogens (*Pn* = P, Sb, Bi) or transition metals which replace Fe (*TM* = Cr, Co, Ni, Ru, Rh, Ir *etc*) are synthesized and examined, see review [6].

All mentioned materials are anisotropic (quasi-two-dimensional) systems with a crystal structure including the negatively charged layers $[TM_2Pn_2]^{\delta-}$ alternating with the positively charged atomic sheets or layers. Moreover, superconductivity in all these systems superconductors is attributed to the states of the $[TM_2Pn_2]$ layers which make the decisive contribution to the near-Fermi region of these materials, while the positively charged blocs (for example, $[LnO]^{\delta+}$ or $A^{\delta+}$) serve as the so-called charge reservoir layers [1-5].

The available data [6] make it possible to note [7] an interesting empirical correlation between the space separation of the conducting [FeAs] blocs in the FeAs superconductors and their critical temperatures $T_c$'s. For example, the maximum $T_c$'s (about 55-56K) have been reached to-date for doped "1111" phases: $Gd_{1-x}Th_yFeAsO$ [2] and $SmFeAsO_{1-x}F_x$ [8]. In these materials, the distances $L$ between the neighboring [FeAs]/[FeAs] blocs separated by $[LnO]^{\delta+}$ layers are about 8.7 Å. For the "122" phases where the neighboring [FeAs]/[FeAs] blocs are separated by the monoatomic sheets of alkali earth metals ($L$~ 6.5 Å), the $T_c$'s are reduced down to 38K. Even lower $T_c$'s ~18 K were revealed for the "111" phases where the [FeAs]/[FeAs] layers are separated by the alkali metal sheets ($L$~ 6.4 Å). Let us also mention the related iron-containing SC FeSe with a tetragonal structure (anti-PbO type) and directly-contacting [FeSe]/[FeSe] layers, for which $T_c$ under normal conditions does not exceed 7-8K, see [9]. The nature of this correlation is still unclear. One of the possible explanations [7] is based on the hypothesis that spin fluctuations are important in the superconductivity mechanism of the FeAs materials when the increasing space separation of the superconducting [FeAs]/[FeAs] blocks hinders the formation of the long-range antiferromagnetic order and, thus, promotes an increase in $T_c$'s.

In view of these circumstances, the recent report [10] on the synthesis of a new five-component phase $Sr_4Sc_2Fe_2P_2O_6$ with $T_c$ ~17K is very interesting. Really, this material adopts an alternating stacking of $[Fe_2P_2]$ and perovskite-based $[Sr_4Sc_2O_6]$ blocks, where $[Fe_2P_2]/[Fe_2P_2]$ separation ($L$~ 15.5 Å) is the longest, and the critical temperature is the highest among all of known FeP SCs, see [11]. In addition, the high $T_c$ for $Sr_4Sc_2Fe_2P_2O_6$ may be partly due to the higher tetrahedral symmetry at FeP layer than in other iron phosphide-oxides such as LaFePO.

In this Communication, we present the first results of the band structure calculations of a new layered 17K SC $Sr_4Sc_2Fe_2P_2O_6$. In result, for this phase the optimized structural data, the energy bands, total and partial densities of states, Fermi surface topology, low-temperature electron specific heat and molar Pauli paramagnetic susceptibility have been determined and discussed in comparison with hypothetical isostructural iron arsenide-oxide phase $Sr_4Sc_2Fe_2As_2O_6$ and related layered FeAs and FeP superconductors.

## Computational details

Our calculations were carried out by means of the full-potential method with mixed basis APW+lo (FLAPW) implemented in the WIEN2k suite of programs [12]. The generalized gradient approximation (GGA) to exchange-correlation potential in the PBE form [13] was used. The plane-wave expansion was taken up to $R_{MT} \times K_{MAX}$ equal to 9, and the $k$ sampling with 12×12×3 $k$-points in the Brillouin zone was used. The calculations were performed with full-lattice optimizations including the atomic positions. The self-consistent calculations were considered to be converged when the difference in the total energy of the crystal did not exceed 0.1 mRy and the difference in the total electronic charge did not exceed 0.001 $e$ as calculated at consecutive steps.

## Results and discussion

### *1. Structural properties.*

The newly synthesized $Sr_4Sc_2Fe_2P_2O_6$ adopts complicated tetragonal (space group *P4/nmm*) crystal structure; the experimental obtained lattice constants are $a$ = 4.016 Å and $c$ = 15.543 Å [10]. This material consists of stacking of anti-fluorite $[Fe_2P_2]$ blocks and perovskite-like $[Sr_4Sc_2O_6]$ blocks as depicted in Fig. 1.

As the detailed atomic coordinates for $Sr_4Sc_2Fe_2P_2O_6$ are absent [10], at the first stage the complete structural optimization of this phase was performed over both the lattice parameters and the atomic positions (see Table 1). The calculated theoretical XRD pattern of $Sr_4Sc_2Fe_2P_2O_6$ phase coincides well with powder XRD data [10], see

Fig. 2. Besides, the calculated lattice constants for $Sr_4Sc_2Fe_2P_2O_6$ phase ($a^{calc}$ = 4.008 Å and $c^{calc}$ = 15.444 Å) are in reasonable agreement with the available experiment [10]: the divergences $(a^{calc} - a^{exp})/a^{exp}$ and $(c^{calc} - c^{exp})/c^{exp}$ are -0.002 and -0.006, respectively, and these divergences should be attributed probably to presence of some amount (~ 10%) of secondary phase $SrFe_2P_2$ in the synthesized samples [10]. According to our calculations, the Fe-P-Fe angles are 74.8° and 125° and the Fe-Fe and Fe-P distances are 2.82 Å and 2.25 Å, respectively. The corresponding experimental values [10] are: 74.6°, 115.0° and 2.84 Å.

For hypothetical iron arsenide-oxide phase $Sr_4Sc_2Fe_2As_2O_6$ the calculated lattice constants ($a^{calc}$ = 4.036 Å and $c^{calc}$ = 15.534 Å) are higher than for $Sr_4Sc_2Fe_2P_2O_6$ - as should be expected at replacement of small phosphorus atoms (with the atomic radius $R^{at}$ = 1.30 Å) for larger arsenic atoms ($R^{at}$ = 1.48 Å); the simultaneous growth of Fe-Fe (2.85 Å) and Fe-As (2.35 Å) distances occurs. At the same time our results show that replacement P→As leads to some *anisotropic deformations* of the crystal structure caused by strong anisotropy of inter-atomic bonds, see also [11,14]. Really, the growth of $a^{calc}$ for $Sr_4Sc_2Fe_2As_2O_6$ in comparison with $Sr_4Sc_2Fe_2P_2O_6$ is at about 0.05 Å *versus* the growth of $c^{calc}$ – at about 0.09 Å. Besides, the thickness of the [$Fe_2As_2$] blocks in comparison with [$Fe_2P_2$] blocks has increased, whereas the [$Sr_4Sc_2O_6$] blocks in $Sr_4Sc_2Fe_2As_2O_6$ are compressed as compared with $Sr_4Sc_2Fe_2P_2O_6$ – for example the Sr-Sr distances (between the strontium atoms placed on the opposite sites of blocks [$Sr_4Sc_2O_6$]) for $Sr_4Sc_2Fe_2As_2O_6$ is 3.58 Å – *versus* 3.64 Å for $Sr_4Sc_2Fe_2P_2O_6$. Note also that the calculated Fe-As-Fe angles (74.7° and 118.2°) for $Sr_4Sc_2Fe_2As_2O_6$ as compared to that of $Sr_4Sc_2Fe_2P_2O_6$ are more close to ideal angles of tetrahedra (109.5°), and this can be considered as a factor favorable for superconductivity [10,15].

*2. Electronic band structure and Fermi surface*

Figure 3 shows the band structure of $Sr_4Sc_2Fe_2P_2O_6$ and $Sr_4Sc_2Fe_2As_2O_6$ with optimized geometry as calculated along the high-symmetry *k* lines. For $Sr_4Sc_2Fe_2P_2O_6$ the occupy bands form three main groups in the intervals -11.9 eV ÷ -10.3 eV; -5.8 ÷ -1.4 eV and from- 1.4 eV up to $E_F$. The Fermi level is crossed by low-dispersive bands with mainly Fe 3*d* character; these bands form two electron pockets centered at *M* and two hole pockets centered at Γ. Note that the similar picture of the low-energy band structure (where the bands which crossed the Fermi level are mainly of Fe $d_{xy,xz,yz}$ symmetry) was established for LaFeAsO [16], whereas a significant difference between $Sr_4Sc_2Fe_2P_2O_6$ and the related low-temperature SC ($T_c$~2-5K) LaFePO [16,17] comes from the third pocket (formed basically by Fe $d_{3z^2-r^2}$ band) centered along the Γ→*Z* direction for the latter. This difference in the low-energy band structure is mainly due to its great sensitivity to the inter-atomic distances and bonding angles in [$Fe_2Pn_2$] blocks; when these parameters for $Sr_4Sc_2Fe_2P_2O_6$ are more close to LaFeAsO than to LaFePO, see also [10,14]. Note also that the low-energy Fe 3$d$-like bands for recently synthesized phase $(Sr_3Sc_2O_5)Fe_2As_2$ (where the conducting [FeAs]/[FeAs] blocks are also spaced by a very long distance: ~ 13.4 Å) are placed higher (over the Fermi level [18]) than for $Sr_4Sc_2Fe_2P_2O_6$, and this fact may be related to the absence of superconductivity for ideal $(Sr_3Sc_2O_5)Fe_2As_2$, which probably will become SC at the electron doping.

Comparison of the data for $Sr_4Sc_2Fe_2P_2O_6$ and hypothetical $Sr_4Sc_2Fe_2As_2O_6$ (Fig. 3) shows that these phases preserve the common features of their band structures; the obvious difference is the width and composition of the main band groups, see Table 2.

The Fermi surface (FS) of $Sr_4Sc_2Fe_2P_2O_6$ is made of five sheets (five bands are crossing the Fermi level as shown by Fig. 4), which are cylindrical-like and parallel to the $k_z$ direction. Three of them are hole-like and are centered along the Γ−Z direction while the two others (electronic-like) are centered along the A−M direction line.

For $Sr_4Sc_2Fe_2As_2O_6$ the FS is made of four cylindrical-like sheets, among them two sheets are hole-like and are centered along the Γ−Z direction while the others (electronic-like) are centered along the A−M line. All sheets are parallel to the $k_z$ direction. As compared with FS of $Sr_4Sc_2Fe_2P_2O_6$ for $Sr_4Sc_2Fe_2As_2O_6$ the sheets of FS are "compressed'.

*3. Density of states*

For further description of the electronic spectra of $Sr_4Sc_2Fe_2P_2O_6$ and $Sr_4Sc_2Fe_2As_2O_6$ we have plotted in Fig. 5 the total and partial densities of states (DOSs). For $Sr_4Sc_2Fe_2P_2O_6$ at high binding energy (in the interval from -11.9 eV up to -10.3 eV), the DOS is almost completely made from P 3$s$ orbitals - with very small admixture of Sr 5$p$ orbitals. Here we shall note, that the contributions from the valence s,$p$ states of Sr to the all occupied bands are quite negligible, *i.e.* in $Sr_4Sc_2Fe_2P_2O_6$ these atoms are in the form of cations close to $Sr^{2+}$.

In the next interval from -5.8 eV up to -1.4 eV all the atoms of $Sr_4Sc_2Fe_2P_2O_6$ phase are contributing to the DOS and these states are responsible for the hybridization effects, i.e. for inter-atomic covalent bonding. Taking into account the distribution of the corresponding atoms over the $[Fe_2P_2]$ and $[Sr_4Sc_2O_6]$ blocks (see Fig. 1) it points to the formation of the Fe–P and Sc–O covalent bonds due to the hybridization of Fe 3$d$ - P 3$p$ states and Sc 3$d$ - O 2$p$ states, respectively, see alsi below.

For $Sr_4Sc_2Fe_2As_2O_6$ the bands with high binding energies are mainly of As 4$p$ character, whereas the states, placed in the interval from -5.3 eV up to -2.1 eV, form the Fe–As and Sc–O bonds owing to hybridization of Fe 3$d$ - As 4$p$ and Sc 3$d$ - O 2$p$ states.

As electrons near the Fermi surface are involved in formation of superconducting state, it is important to figure out their nature. It is seen (Fig. 5) that the regions of DOSs near the Fermi level in $Sr_4Sc_2Fe_2P_2O_6$ and $Sr_4Sc_2Fe_2As_2O_6$ are formed exclusively by states of $[Fe_2P_2]$ and $[Fe_2As_2]$ blocks, respectively. Therefore, the conduction in these phasees is expected to be strongly anisotropic, i.e. happening mainly in these blocks. Notice that the contributions from P (As) states are much smaller than the contributions from Fe 3*d* orbitals, see also Table 3, where the total and orbital decomposed partial DOSs at the Fermi level, $N(E_F)$, are shown.

For $Sr_4Sc_2Fe_2As_2O_6$ the $N(E_F)$ value is at about 5% lower than for $Sr_4Sc_2Fe_2P_2O_6$. This effect may be attributed mainly to the lowering of the contributions of As states in $N(E_F)$, whereas the density of Fe 3*d* states at the Fermi level has increased at about 17%, Table 3. Note also that for $Sr_4Sc_2Fe_2P_2O_6$ the Fermi level is located on flat plateau of DOS, whereas for $Sr_4Sc_2Fe_2As_2O_6$ - on an abrupt slope, Fig. 5. Thus the hole doping will lead to more significant growth of $N(E_F)$ for $Sr_4Sc_2Fe_2As_2O_6$ than for $Sr_4Sc_2Fe_2P_2O_6$. For both phases the electron doping will be accompanied by the lowering of $N(E_F)$.

The obtained data allow us also to estimate the Sommerfeld constants (γ) and the Pauli paramagnetic susceptibility (χ) for examined phases under assumption of the free electron model as $\gamma = (\pi^2/3)N(E_F)k^2_B$ and $\chi = \mu_B^2 N(E_F)$. It is seen from Table 3 that both γ and χ decrease slightly at replacement of phosphorus by arsenic. On the other hand, the values of γ and χ obtained for $Sr_4Sc_2Fe_2P_2O_6$ and $Sr_4Sc_2Fe_2As_2O_6$ are comparable with the same for others known FeAs and FeP SCs – for example $\gamma \sim 3.7$ $mJ{\cdot}K^{-2}{\cdot}mol^{-1}$ for $LaFeAsO_{1-x}F_x$, ~ 10 $mJ{\cdot}K^{-2}{\cdot}mol^{-1}$ for LaFePO or ~10.8 $mJ{\cdot}K^{-2}{\cdot}mol^{-1}$ for $BaNi_2As_2$, see [19].

## *4. Chemical bonding*

To describe the inter-atomic bonding for new phase $Sr_4Sc_2Fe_2P_2O_6$, we begin with a simple ionic picture, which considers the standard oxidation numbers of atoms: $Sr^{2+}$, $Sc^{3+}$, $Fe^{2+}$, $P^{3-}$ and $O^{2-}$. Taking into account the distributions of atoms in above mentioned blocks of $Sr_4Sc_2Fe_2P_2O_6$, the ionic formula of this phase can be presented as $[(Fe^{2+})_2(P^{3-})_2]^{2-}[(Sr^{2+})_4(Sc^{3+})_2(O^{2-})_6]^{2+}$. Hence, in the $Sr_4Sc_2Fe_2P_2O_6$, the charge transfer occurs from positively charged blocks $[Sr_4Sc_2O_6]^{\delta+}$ to negatively charged conducting blocks $[Fe_2P_2]^{\delta-}$ (as well as for the groups of the others "1111", "122" and "111" FeAs and FeP superconductors, see [6,11,17,18,20,21]), and between these blocks the ionic bonding takes place. This picture is clearly visible in Fig. 6, where the charge density map for $Sr_4Sc_2Fe_2P_2O_6$ is depicted. Besides, inside each block: $[Sr_4Sc_2O_6]$ and $[Fe_2P_2]$, the ionic bonding takes place between the ions with opposite charges: $(Sr^{2+},Sc^{3+})$ - $O^{2-}$ and $Fe^{2+}$ - $P^{3-}$. Further, the charge density distribution (Fig. 6) reveals also the mentioned covalent bonding Fe-P and Sc-O inside $[Fe_2P_2]$ and $[Sr_4Sc_2O_6]$ blocks, respectively. In addition inside $[Fe_2P_2]$ blocks the metallic-like Fe-Fe bonding occurs due to overlapping of near-Fermi Fe-3*d* states, see Fig. 4.

Thus, summarizing the above results, the intra-atomic bonding for $Sr_4Sc_2Fe_2P_2O_6$, as well as for $Sr_4Sc_2Fe_2As_2O_6$, phases can be classified as high-anisotropic mixture of ionic, covalent and metallic contributions.

## Conclusions

In summary, we present the first *ab initio* study of the structural and electronic properties of the newly synthesized tetragonal iron phosphide-oxide: 17K superconductor $Sr_4Sc_2Fe_2P_2O_6$ in comparison with hypothetical isostructural iron arsenide-oxide phase $Sr_4Sc_2Fe_2As_2O_6$.

For both phases we obtain almost the same band structure picture around the Fermi level, where the Fermi level is crossed by low-dispersive two-dimensional-like bands with mainly Fe 3*d* character. The Fermi surfaces of $Sr_4Sc_2Fe_2P_2O_6$ and $Sr_4Sc_2Fe_2As_2O_6$ are also very similar each other and for others FeAs and FeP SCs. Thus the hypothetical phase $Sr_4Sc_2Fe_2As_2O_6$ may be expected as a possible superconductor, and it will be of interest to synthesize and experimentally probe this phase to search for superconductivity. Also, the further theoretical and experimental studies are necessary to clarify the possible effects of hole – or electron doping on the properties of these interesting materials.

**Table 1.**
The optimized lattice parameters (*a* and *c*, in Å) and atomic positions for $Sr_4Sc_2Fe_2P_2O_6$ and $Sr_4Sc_2Fe_2As_2O_6$.

| system | $Sr_4Sc_2O_6Fe_2P_2$ | | | $Sr_4Sc_2O_6Fe_2As_2$ | | |
|---|---|---|---|---|---|---|
| *a* | 4.008 | | | 4.036 | | |
| *c* | 15.444 | | | 15.534 | | |
| $Sr_1$ (2c) | 0.25 | 0.25 | 0.3232 | 0.25 | 0.25 | 0.3176 |
| $Sr_2$ (2c) | 0.25 | 0.25 | 0.0878 | 0.25 | 0.25 | 0.0870 |
| Sc(2c) | 0.25 | 0.25 | 0.8024 | 0.25 | 0.25 | 0.8042 |
| $O_1$ (4f) | 0.75 | 0.25 | 0.2213 | 0.75 | 0.25 | 0.2179 |
| $O_2$ (2c) | 0.25 | 0.25 | 0.9319 | 0.25 | 0.25 | 0.9325 |
| Fe (2b) | 0.75 | 0.25 | 0.5 | 0.75 | 0.25 | 0.5 |
| P(As) (2c) | 0.25 | 0.25 | 0.5673 | 0.25 | 0.25 | 0.5778 |

**Table 2.**
Calculated bandwidths (in eV) of occupied states for $Sr_4Sc_2Fe_2P_2O_6$ and $Sr_4Sc_2Fe_2As_2O_6$.

| system | $Sr_4Sc_2Fe_2P_2O_6$ | $Sr_4Sc_2Fe_2As_2O_6$ |
|---|---|---|
| Valence band (up to $E_F$) | 5.8 | 5.3 |
| Bang gap | 4.5 | 5.5 |
| Quasi-core pnictogen *s* band | 1.6 | 1.3 |

**Table 3.**

Total and partial densities of states at the Fermi level (in states/eV·form.unit), electronic heat capacity $\gamma$ (in $mJ \cdot K^{-2} \cdot mol^{-1}$) and molar Pauli paramagnetic susceptibility $\chi$ (in $10^{-4}$ emu/mol) for $Sr_4Sc_2Fe_2P_2O_6$ and $Sr_4Sc_2Fe_2As_2O_6$.

| system | $N^{Fe3d}(E_F)$ | $N^{(P,As)p}(E_F)$ | $N^{tot}(E_F)$ | $\gamma$ | $\chi$ |
|---|---|---|---|---|---|
| $Sr_4Sc_2Fe_2P_2O_6$ | 2.378 | 0.209 | 3.954 | 9.32 | 1.27 |
| $Sr_4Sc_2Fe_2As_2O_6$ | 2.779 | 0.081 | 3.754 | 8.85 | 1.21 |

## FIGURES

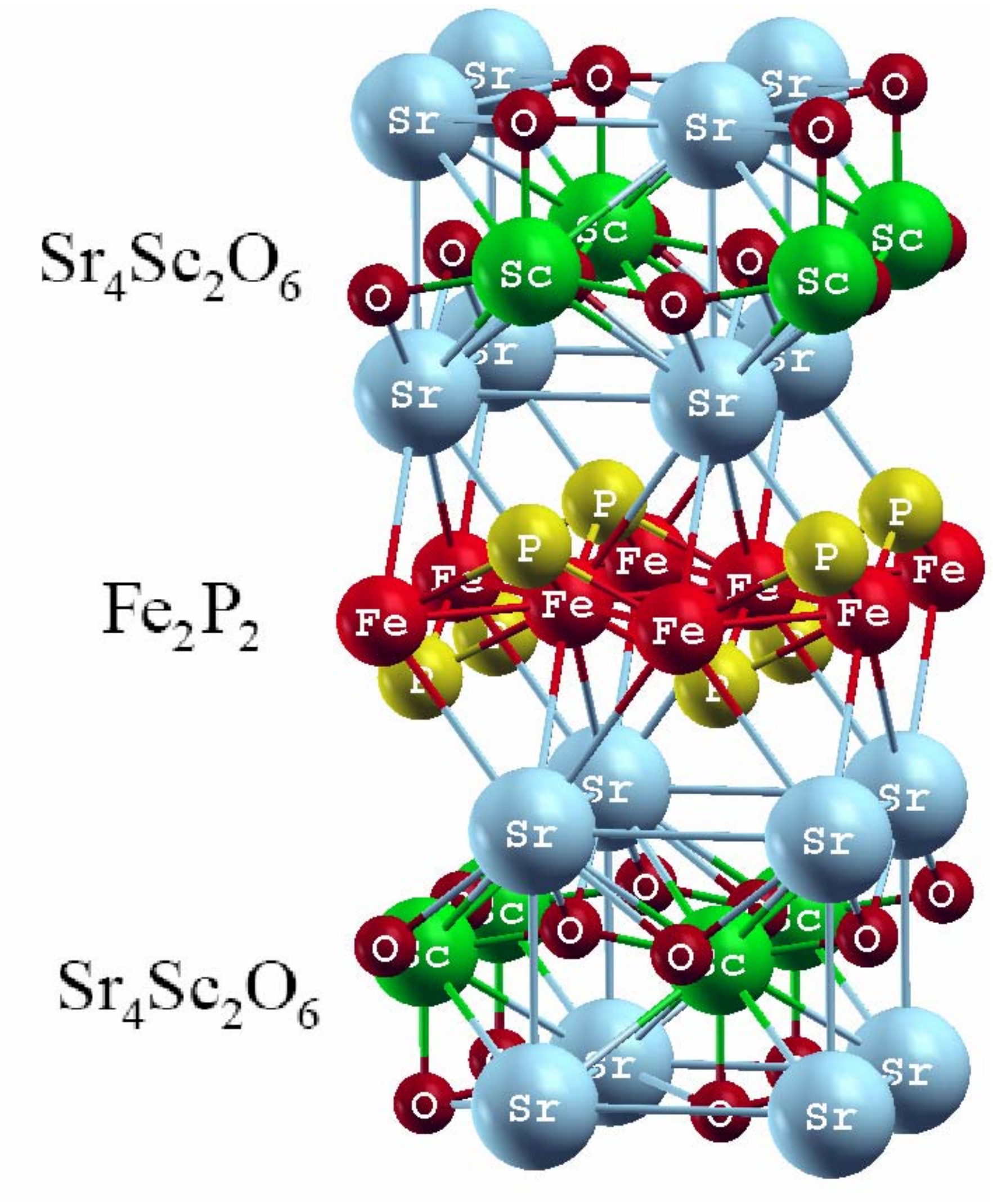

**Fig. 1.** Crystal structure of the phase $Sr_4Sc_2Fe_2P_2O_6$ (space group *P4/nmm*). The main building blocks [$Fe_2P_2$] and [$Sr_4Sc_2O_6$] are shown.

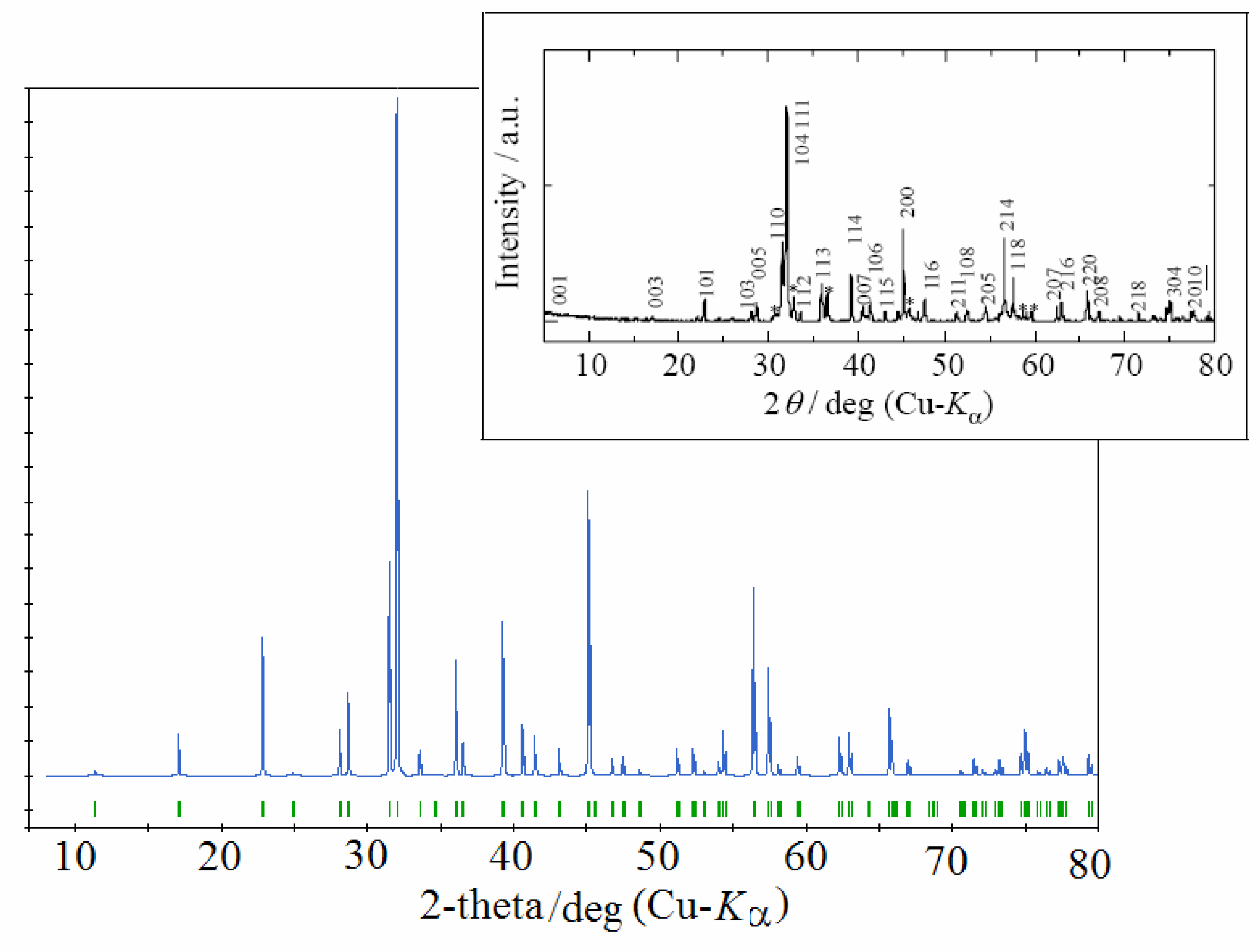

**Fig. 2.** Calculated theoretical XRD pattern of $Sr_4Sc_2Fe_2P_2O_6$ phase as compared with powder XRD pattern (*on an insert* [10]).

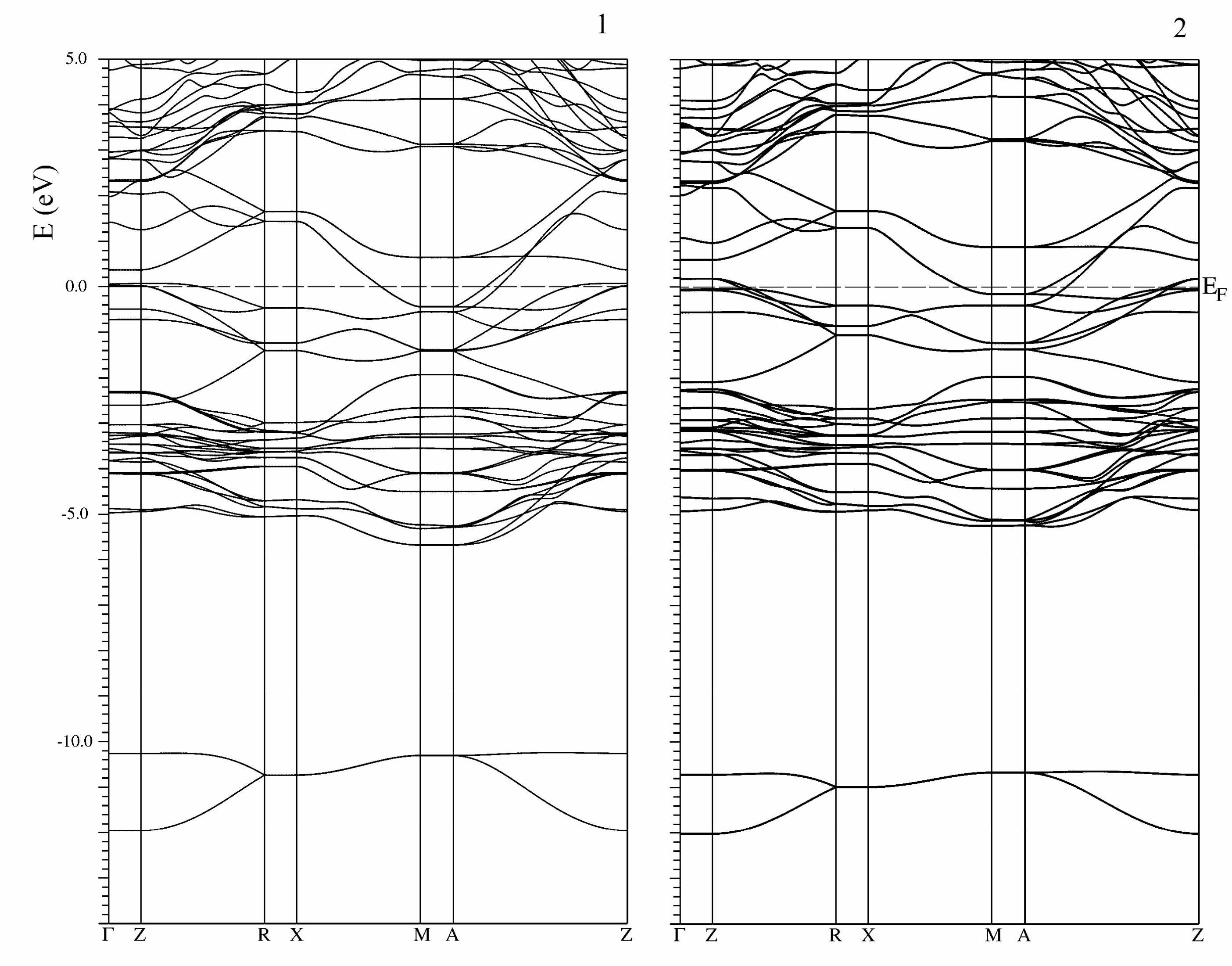

**Fig. 3.** Electronic bands for $Sr_4Sc_2Fe_2P_2O_6$ (1) and $Sr_4Sc_2Fe_2As_2O_6$ (2)

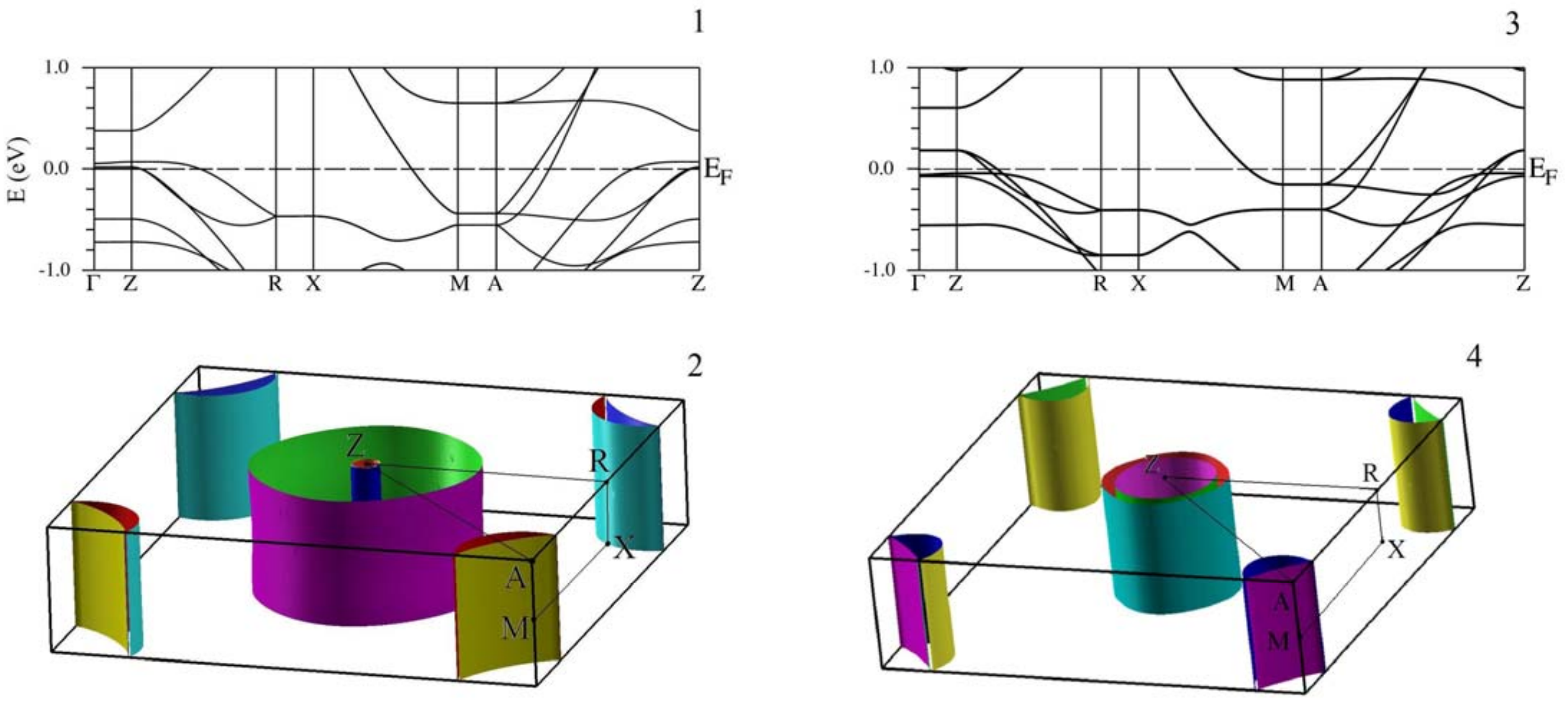

**Fig. 4.** The near-Fermi bands and Fermi surfaces for $Sr_4Sc_2Fe_2P_2O_6$ (1,2) and $Sr_4Sc_2Fe_2As_2O_6$ (3,4)

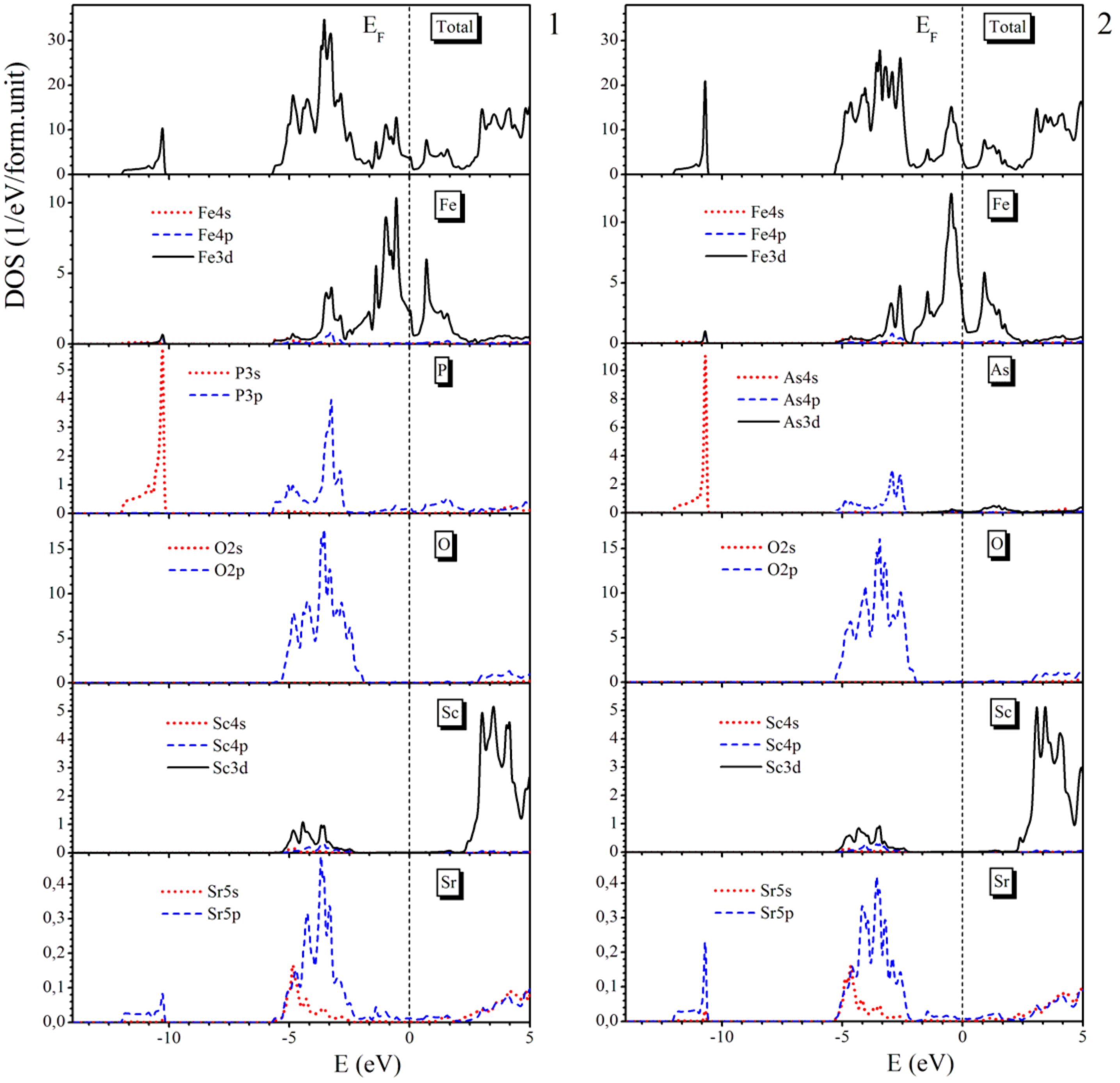

**Fig. 5.** Total (*upper panels*) and partial densities of states (*bottom panels*) for $Sr_4Sc_2Fe_2P_2O_6$ (1) and $Sr_4Sc_2Fe_2As_2O_6$ (2).

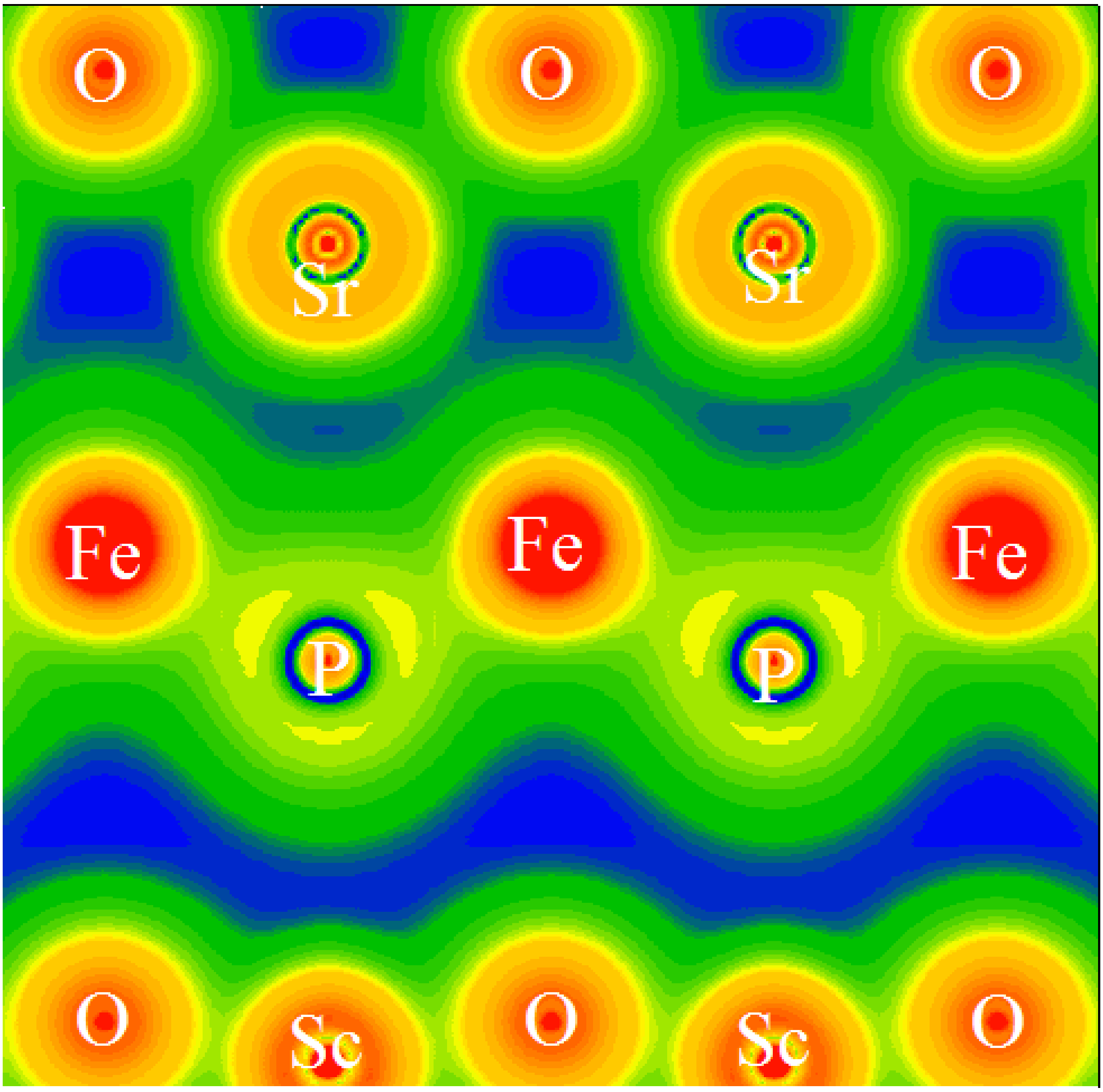

**Fig. 6.** Valence charge density map (in e/Å$^3$) for $Sr_4Sc_2Fe_2P_2O_6$.